\documentclass[preprint,prd,aps,showpacs,showkeys,nofootinbib]{revtex4}
\usepackage{graphicx}
\usepackage{dcolumn}
\usepackage{bm}
\begin{document}

\title{Top Quark Forward-Backward Asymmetry in the \\Little Higgs Model}
\author{Xing-Dao Guo$^{1}$, Yin-Jie Zhang$^{1}$, Shu-Min Zhao$^{1\footnotemark[1]}$, Tai-Fu Feng$^{1\footnotemark[2]}$, Xu-Hao Yuan$^{2}$, Xue-Qian Li$^3$}
 \affiliation{\small{$^1$ Department of Physics and Technology, Hebei University, Baoding 071002,
 China} \\ \small{$^2$ Center for High Energy Physics, Department of Engineering Physics, Tsinghua University, Beijing 100084,
 China}\\ \small{$^3$ Department of Physics, Nankai University, Tianjin 300071,
 China}}
\footnotetext[1]{zhaosm@mail.hbu.edu.cn}
\footnotetext[2]{fengtf@dlut.edu.cn}
\date{\today}
\begin{abstract}
We calculate the forward backward asymmetry of the top-pair production at TEVATRON up to next to leading order (NLO) in the little Higgs model (LHM).  We find that the contribution of $Z_H$ can be large enough to make up the gap between standard model (SM) prediction and data. With the database of $7.65\pm0.20\pm0.36$ pb, therefore, the parameter space for flavor-changing coupling of $Z_H$ is constrained. Thus this model can result in the required asymmetry while the total cross section of top-pair production remaining consistent with data.
\end{abstract}


\maketitle

\section{Introduction}
With the new discovery of  the 125 GeV boson made by  the ATLAS and CMS collaborations one is tempted to consider that the particle zoo predicted by the standard model is indeed complete. However, the story is not over yet! There are still several notable discrepancies between the SM predictions and data, therefore the new physics beyond SM (BSM) must exist, but what is it? In fact, many such BSM models are available at present, but we do not know which one or ones are the right one or none of them. The strategy is to apply such models to study the phenomena observed in high energy experiments, and then compare the theoretical predictions with the data to confirm or negate the model, at least constrain the concerned parameter space for the model.

Because of its heavy mass which is supposed to be closer to the scale of possible new physics, the processes involving top quark are more easily affected by the new physics.   At TEVATRON of the Fermilab  top quark production and decays have been investigated theoretically and experimentally\cite{Willenbrock:2000vv,Peters:2011an}.  Meanwhile some discrepancies between theoretical predictions (mainly SM) for top processes and data were observed which encourage exploration of BSM.  One of the discrepancies is the large forward-backward(FB) asymmetry (AFB). The asymmetry parameter is defined as
\begin{equation}
A_{FB}=\frac{N_t(\cos\theta>0)-N_t(\cos\theta<0)}{N_t(\cos\theta>0)+N_t(\cos\theta<0)}\label{fb},
\end{equation}
where $\theta$ is the angle spanned between the outgoing top quark and incoming proton beam.
The measurements of the CDF and D0 Collaborations yield $A_{FB}=0.158\pm 0.075$\cite{Aaltonen:2011kc}, $A_{FB}=0.162\pm 0.047$\cite{Aaltonen:2012CDFnote} and $A_{FB}=0.196\pm 0.065$\cite{Abazov:2011prd}, and the central values are significantly larger than the Standard Model(SM) prediction, $A_{FB}^{SM}=0.089$\cite{Hollik:2011ps}.  This discrepancy has motivated people to consider additional contributions from  new physics BSM\cite{sj}.

In some models  an exchange of new particle of color-octet at s-channel contributes\cite{phf}. Instead, a color singlet particle which mediates the interaction  may also provide a substantial contribution if it has both vector and axial vector coupling to $q\bar{q}$ and $t\bar{t}$\cite{kc}. This kind of models receive severe constraints from the data of the $t\bar t$ production rate. However, as one notices, the tree diagram of this kind of models may interfere with the QCD box diagrams and results in a substantial contribution to the  top forward-backward asymmetry. The little Higgs model (LHM) is just one of such models.

The LHM can solve the hierarchy problem of particle physics, so is a favorable extension of the SM. The LHM begins with an $SU(5)$ global symmetry, which is spontaneously broken down to its subgroup $SO(5)$ via a non-zero vacuum expectation value $f$, leaving 14 Goldstone bosons which transform under the electroweak gauge group as a real singlet $1_0$, a real triplet $3_0$, a complex doublet $2_{\pm \frac{1}{2}}$, and a complex triplet $3_{\pm1}$. The real singlet and the triplet become the longitudinal components of the gauge bosons associated with the broken gauge groups, giving them masses of the order $f$. These gauge bosons are ($w^\pm_L$, $A_L$, $Z_L$, $w^\pm_H$, $A_H$ and $Z_H$ ). $w^\pm_H$, $A_H$ and $Z_H$ are new particles in the LHM.

The difference of the rapidities of outgoing $Q$ and $\bar Q$ is directly related to $\theta$ as  \cite{kc}
\begin{equation}
y_t-y_{\bar{t}}=2{\rm arctanh}\Big(\sqrt{1-\frac{4m_t^2}{\hat{s}}}\cos\theta\Big),\label{rap}
\end{equation}
where $\hat{s}=(p_t+p_{\bar t})^2$ and the rapidity is defined as
\begin{equation}
y_t=\frac{1}{2}\ln[\frac{E_t+p_t}{E_t-p_t}],
\end{equation}
and $E_t$ and $p_t$ stand for the energy and longitudinal component of the momentum of the top quark respectively. Obviously, the rapidity difference is Lorentz invariant.  The sign of $\bigtriangleup y=y_t-y_{\bar{t}}$ is the same as $\cos\theta$, so that the asymmetry in Eq.(\ref{afb}) can be re-defined as
\begin{equation}
A_{fb}\equiv\frac{N_t(y_t-y_{\bar{t}}>0)-N_t(y_t-y_{\bar{t}}<0)}{N_t(y_t-y_{\bar{t}}>0)+N_t(y_t-y_{\bar{t}}<0)}.\label{afb}
\end{equation}
In this work, we are using this definition to account the asymmetry.

In this work we calculate  the top forward-backward asymmetry in the LHM and by comparing the theoretical prediction with the data, we set constraints on the model parameters. We find out that there exists a possible parameter space for the LHM with which the forward backward asymmetry reported by the CDF and D0 collaborations \cite{ta} can be well explained.

This paper is organized as follows. After this introduction, in Section 2, we derive the theoretical formulas for the cross section of the top-pair production
up to leading order in the LHM. In Section 3, a detailed analysis on the asymmetry is presented. The numerical results along with all the input parameters are outlined in Section 4. The last section is devoted to a discussion and our conclusion.

\section{the contribution of leading-order diagram to the asymmetry}
\begin{figure}
\setlength{\unitlength}{1mm}
\begin{center}
\hspace*{0cm}
\includegraphics[width=10cm]{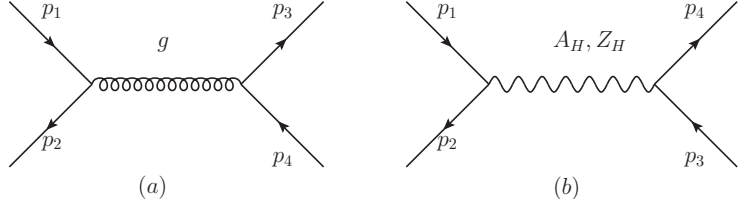}
\vspace{0cm}
\caption{The tree diagrams for the process of $u\bar{u}\rightarrow t\bar{t}$.} \label{tree}
\end{center}
\end{figure}

The LHM consists of four new bosons, but only $A_H$ and $Z_H$ whose masses are respectively $m_{A_H}\propto f g_{vu}$ and $m_{Z_H}\propto f (g_{vu}'+\frac{1}{g_{vu}'})$, can contribute asymmetry through s-channel with couplings
\begin{equation}\begin{array}{rl}
\mathcal{L}_{A_H}=A_H\bar{t}(g_{vt}+g_{at}\gamma^{5})\gamma^{\mu}t+A_H\bar{u}(g_{vu}+g_{au}\gamma^{5})\gamma^{\mu}u,
\end{array}\end{equation}
and
\begin{equation}\begin{array}{rl}
\mathcal{L}_{Z_H}=Z_H\bar{t}(g_{vt}'+g_{at}'\gamma^{5})\gamma^{\mu}t+Z_H\bar{u}(g_{vu}'+g_{au}'\gamma^{5})\gamma^{\mu}u.
\end{array}\end{equation}
The relations among the coupling constants are listed in the appendix of reference\cite{th} and their numerical values are presented in section 4.
An $A^t_{FB}$ which may be consistent with the TEVATRON data, can be generated in $u\bar{u}\to t\bar{t}$ within this model. In order to formulate the whole contribution to the forward backward asymmetry for top quark from LHM, we compute the tree diagrams analytically, and calculate the corresponding box diagrams numerically. The leading-order diagrams are shown in Fig.\ref{tree}.

The amplitude of the first diagram of Fig.\ref{tree} can be written as:
\begin{equation}\begin{array}{rl}
\mathcal{M}_1=\bar{u}(p_4)(-i\gamma^{\mu}g)v(p_3)\frac{-i}{(p_1+p_2)^2}\bar{v}(p_2)(-i\gamma_{\mu}g)u(p_1),
\end{array}\end{equation}
where $p_1$ and $p_2$ respectively stand for the four-momenta of the initial
state ($u\bar{u}$), and $p_{3}$, $p_{4}$ denote the four-momenta of the
final state ($t\bar{t}$).  $p_{1}+p_2=p_3+p_{4}$ stand for the energy-momentum conservation. For the rest two
diagrams, the amplitudes are similar, so we skip them. The contributions at
tree level are shown above, and the numerical results will be given in section 4.

\section{next-to-leading-order diagrams in LHM}

\begin{figure}
\setlength{\unitlength}{1mm}
\begin{center}
\includegraphics[width=9cm]{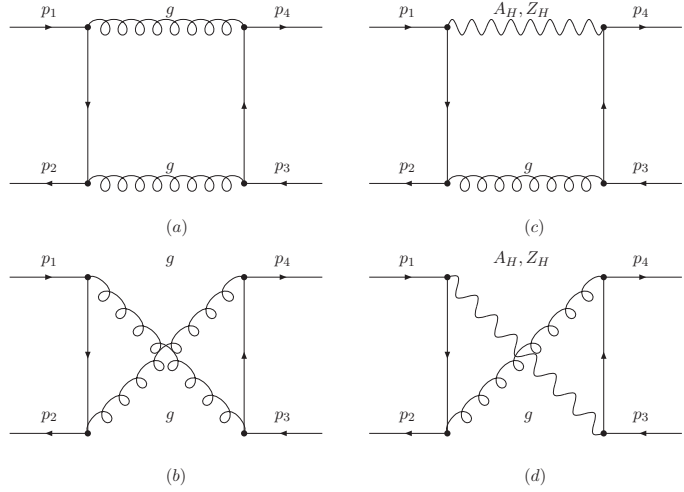}
\vspace{-0cm}
\caption{The box diagrams for the process of $u\bar{u}\rightarrow t\bar{t}$.} \label{box}
\end{center}
\end{figure}

\begin{figure}
\setlength{\unitlength}{1mm}
\begin{center}
\includegraphics[width=9cm]{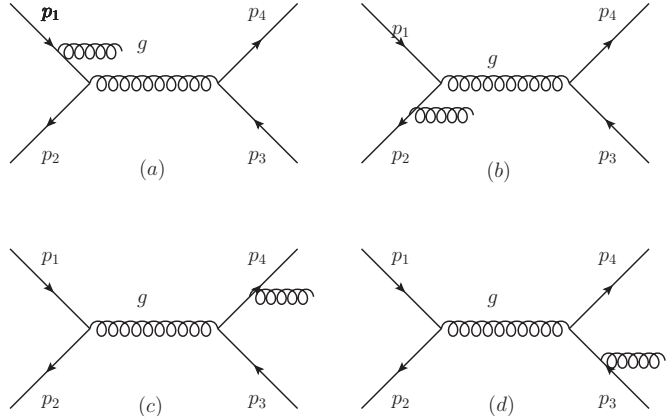}
\vspace{-0cm}
\caption{The real-gluon emission diagrams for the process of $u\bar{u}\rightarrow t\bar{t}+g$ with gluon being the intermediate boson at s-channel. In (a) and (b)
the real gluon is emitted from the initial state while in (c) and (d) it is emitted from one of the produced top quarks. } \label{infrared}
\end{center}
\end{figure}

\begin{figure}
\setlength{\unitlength}{1mm}
\begin{center}
\includegraphics[width=9cm]{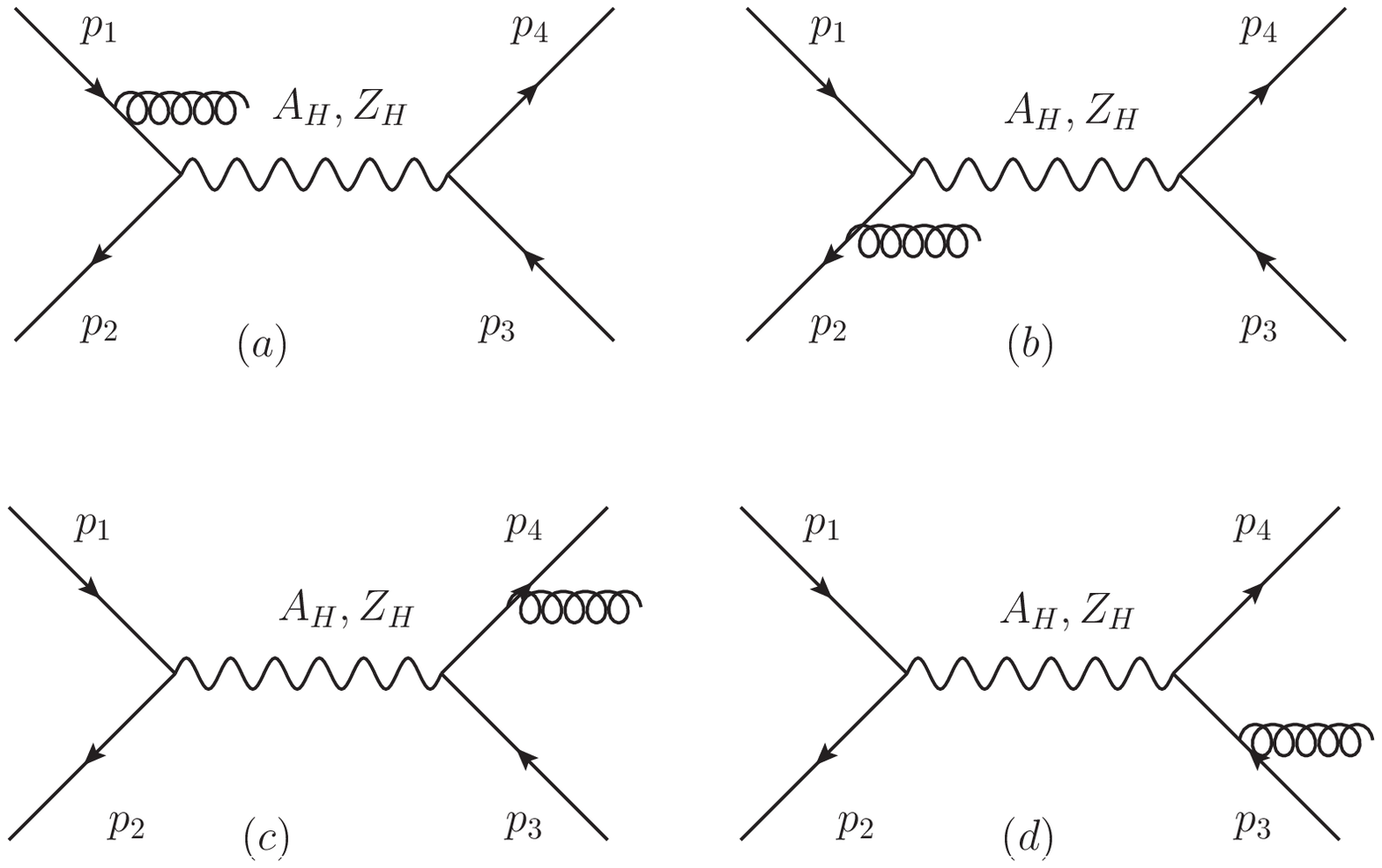}
\vspace{-0cm}
\caption{The The real-gluon emission diagrams for the process of $u\bar{u}\rightarrow t\bar{t}+g$ with $A_H$ or $Z_H$ being the intermediate boson at s-channel. In (a) and (b)
the real gluon is emitted from the initial state while in (c) and (d) it is emitted from one of the produced top quarks.} \label{infrared1}
\end{center}
\end{figure}

In this section, we calculate the next to leading order contribution to the forward-backward asymmetry. The box diagrams contributing to the asymmetry  are shown as Fig.\ref{box}.
And the diagrams for real-gluon emission  are presented in Fig.\ref{infrared} and Fig.\ref{infrared1}.

The amplitude of the first diagram in the Fig.\ref{box} is:
\begin{equation}\begin{array}{rl}
\mathcal{M}_2=&\displaystyle\int\frac{d^4k}{(2\pi)^4}\bar{u}_{\beta}(p_4)\frac{-i}{q^2}(-ig_{s}\gamma^{\mu})\frac{i(\rlap /p_4+\rlap /q+m_4)}{(p_4+q)^2-m_4^2}
(-ig_{s}\gamma^{\nu}T_{\beta\alpha}^{a})v_\alpha(p_3)\\
\times&\bar{v}_{\theta}(p_2)(-ig_{s}\gamma^{\nu}T_{\omega\eta}^{a})\frac{-i}{(p_1+p_2+q)^2}(-ig_{s}\gamma^{\nu})\frac{i(\rlap /p_1+\rlap /q+m_2)}{(p_1+q)^2-m_2^2}u_{\omega}(p_1).
\end{array}\end{equation}

For the rest diagrams, the amplitudes are similar but the coupling vertices are different. The box diagrams have infrared divergences which can be cancelled by adding the real-gluon emission diagrams (Fig.\ref{infrared} and Fig.\ref{infrared1}), and we take into account the interference between initial and final state gluon emission. The dependence of the resultant differential cross section on $\cos\theta$ is:
\begin{equation}
\frac{d{\hat\sigma}}{d \cos\theta}=\frac{2\pi\sqrt{1-\frac{4m_t^2}{s}}}{64\pi^2s}\frac{1}{4}\sum|\mathcal{M}_1+\mathcal{M}_2'|^2
=\frac{\sqrt{1-\frac{4m_t^2}{s}}}{128\pi s}\sum(|\mathcal{M}_1|^2+2Re(\mathcal{M}_1^*\mathcal{M}_2')).
\end{equation}

Then we can obtain the theoretical prediction on the cross section which has been measured for the process by convoluting the sub-processes with the parton distribution functions of proton and anti-proton.
\begin{equation}
\frac{d\sigma_{tot}}{d \cos\theta}=\int_0^1dx\int_0^1dzf(x)f(z)\frac{d{\hat\sigma}}{d \cos\theta}\label{tot}.
\end{equation}
Here we adopt the parton distribution function from CTEQ6M\cite{jp}. The total differential cross section Eq.(\ref{tot}) involves $f(x)$ and $f(z)$ which are the parton distribution functions of proton and anti-proton respectively.
The asymmetry is determined by integrating over the positive and negative range of the $\cos\theta$. In the calculation, we use $y_t-y_{\bar{t}}$ to calculate the asymmetry as given by Eq.(\ref{rap}) and Eq.(\ref{afb}).
The numerical results are presented in section 4.

\section{Numerical results}

In this section, we numerically calculate the differential cross section. We set the mass of top quark as $175$ GeV and neglect the mass of up quark. With the weak-binding approximation, i.e. $ p_q=p_{\bar{q}}$ and $p_q^2=m_q^2$ , we get
\begin{equation}\begin{array}{c}
\displaystyle p_1.p_2=\frac{s}{2},       \displaystyle p_3.p_4=\frac{s}{2}-m_t^2,\\
\displaystyle p_1.p_3=p_2.p_4=\frac{s}{4}\left(1+\sqrt{1-\frac{4m_t^2}{s}}\cos\theta\right),\\
\displaystyle p_1.p_4=p_2.p_3=\frac{s}{4}\left(1-\sqrt{1-\frac{4m_t^2}{s}}\cos\theta\right).
\end{array}\end{equation}
The input parameters which we are going to use in the numerical computations are set as follows  \cite{th,gli,data,wmy,hfr}: $\alpha_s=0.104$ for $\mu=m_t$.
In the LHM\cite{th}, we choose
$$g_{vu}=-0.0292(\frac{3}{a}-2a);\;
g_{au}=-0.0175(\frac{3}{a}-2a);$$
$$g_{vd}=0.2742\frac{3}{a}+0.245a;\;
g_{ad}=0.0175(\frac{3}{a}-2a);$$
$$g_{vt}=-0.0292(\frac{3}{a}-2a)-0.35(\frac{1}{a}+a)b;$$
$$g_{at}=-0.0175(\frac{3}{a}-2a)-0.35(\frac{1}{a}+a)b;$$
$$g_{ve}=0.0525(\frac{3}{a}-2a);\; g_{ae}=0.0175(\frac{3}{a}-2a);$$
and $$m_{A_H}=0.08138(\frac{1}{a}+a)f$$ GeV \cite{th};
where $a$ and $b$
are the parameters in Ref.\cite{th} and  we let them vary
from $0.1$ to $2$ and $0$ to $1$ respectively.
For $Z_H$ boson, $m_{Z_H}=0.0539(36.73g_{u}'+\frac{1}{g_{u}'})f$ GeV, and for simplicity we use
the relation $g_{vu}'=-g_{vd}'=g_{vt}'=-g_{ve}=-g_{au}'=g_{ad}'=g_{at}'=-g_{ae}'$. They could vary from
$-0.0165$ to $-0.33$\cite{th}, and the coupling constant
$\alpha_l=\frac{g_{au}'^2}{4\pi}$ which varies from $0.00002$ to
$0.00867$.
In the calculation, we take $f=500, 1000, 1500$ GeV respectively.

\begin{figure}
\setlength{\unitlength}{1mm}
\begin{center}
\includegraphics[width=12cm]{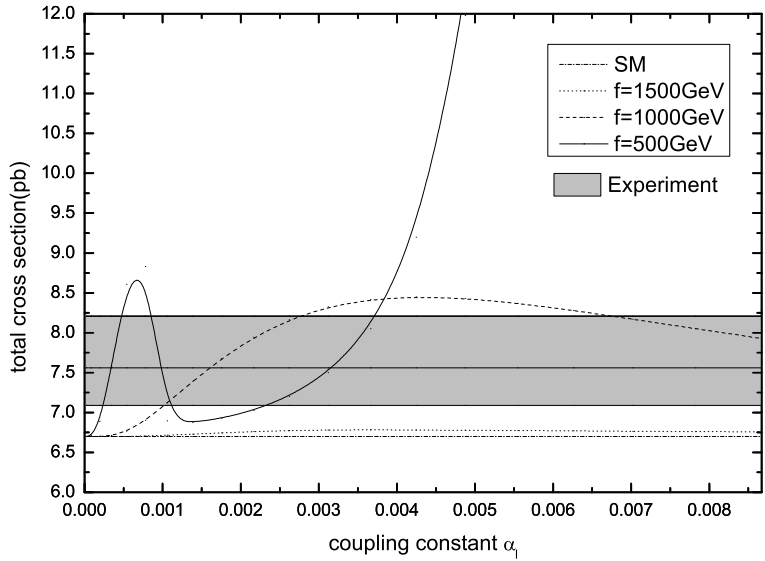}
\caption{The dependence of the total cross section on the coupling constant in the frameworks of SM and LHM with $f=500$GeV(solid), $f=1000$GeV(dashed), $f=1500$GeV(dotted) respectively.
The dash-dotted line stands for the value from SM\cite{Bonciani:1998vc}, and the shadowed region is for experimental data with errors\cite{cdf}} \label{crosssection}
\end{center}
\end{figure}

The important constraint comes from the total production cross section of the $t\bar{t}$ pair measured in recent experiment. The averaged value for $t\bar{t}-$production cross section at Tevatron is \cite{cdf}:
\begin{equation}
\sigma^{exp}(t\bar{t})=7.65\pm0.20\pm0.36 pb.
\end{equation}
The SM prediction on the top quark cross section is $6.7$pb\cite{Bonciani:1998vc} and the AFB was also evaluated within SM \cite{Hollik:2011ps}.
The $A_H$ boson contribution is so small that we can neglect it safely, so $Z_H$
offers the main contribution. We demonstrate the dependence of the total cross section on the coupling constant in Fig.\ref{crosssection}, and dependence of AFB on the coupling constant in Fig.\ref{afb}.

From Fig.\ref{crosssection} we can see that for $f=500$ GeV, when taking the coupling constant as $0.0002\sim0.0005$, $0.0008\sim0.0012$ and $0.0023\sim0.0037$, the theoretical predictions on the total cross section are inside the experimental tolerance range. While for $f=1000$GeV, the theoretical
prediction does not conflict with the experimental data when the coupling constant is about $0.0010\sim0.0028$ and $0.0067\sim0.0087$. As for $f=1500$ GeV, one can scarcely
find a region that can match the experimental data.

\begin{figure}
\setlength{\unitlength}{1mm}
\begin{center}
\includegraphics[width=12cm]{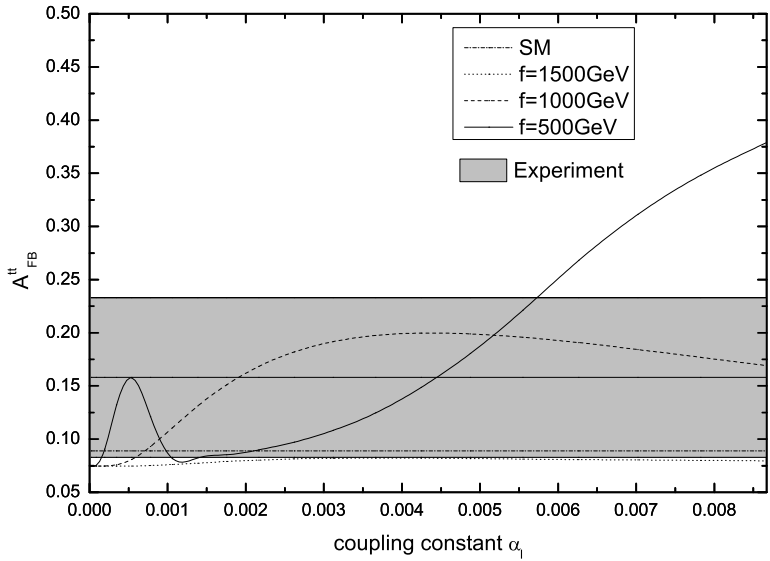}
\caption{The AFB versus coupling constant in the frameworks of SM and LHM with $f=500$GeV(solid), $f=1000$GeV(dashed), $f=1500$GeV(dotted).
The dash-dotted line stands for the value from SM\cite{Bonciani:1998vc}, and the shadowed region is for experimental data with errors\cite{Aaltonen:2011kc}.} \label{afb}
\end{center}
\end{figure}

In Fig.\ref{afb} we find that for $f=500$ GeV, when taking the coupling constant below $0.0057$, the AFB almost matches the experimental data.
for $f=1000$ GeV, the theoretical prediction fits the experimental value when the coupling constant is above $0.0006$. But it is hard
to find an area that can match the experimental data with $f=1500$ GeV.

\begin{figure}
\setlength{\unitlength}{1mm}
\begin{center}
\includegraphics[width=12cm]{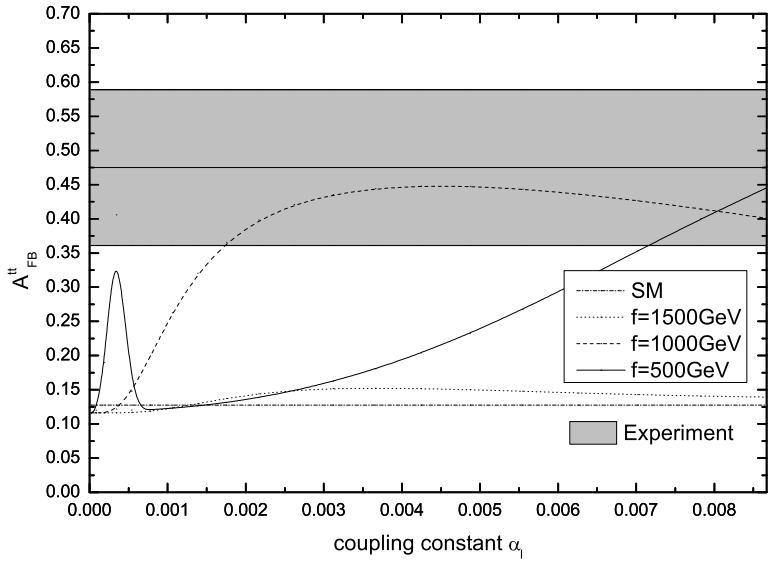}
\caption{The AFB with a cut of $M_{t\bar{t}}>450$GeV versus the coupling constant in the frameworks of SM and LHM with $f=500$ GeV(solid), $f=1000$ GeV(dashed), $f=1500$ GeV(dotted).
The dash-dotted line stands for the value from SM\cite{Hollik:2011ps}, and the shadowed region is for the experimental data\cite{Aaltonen:2011kc}.} \label{mcut}
\end{center}
\end{figure}

\begin{figure}
\setlength{\unitlength}{1mm}
\begin{center}
\includegraphics[width=12cm]{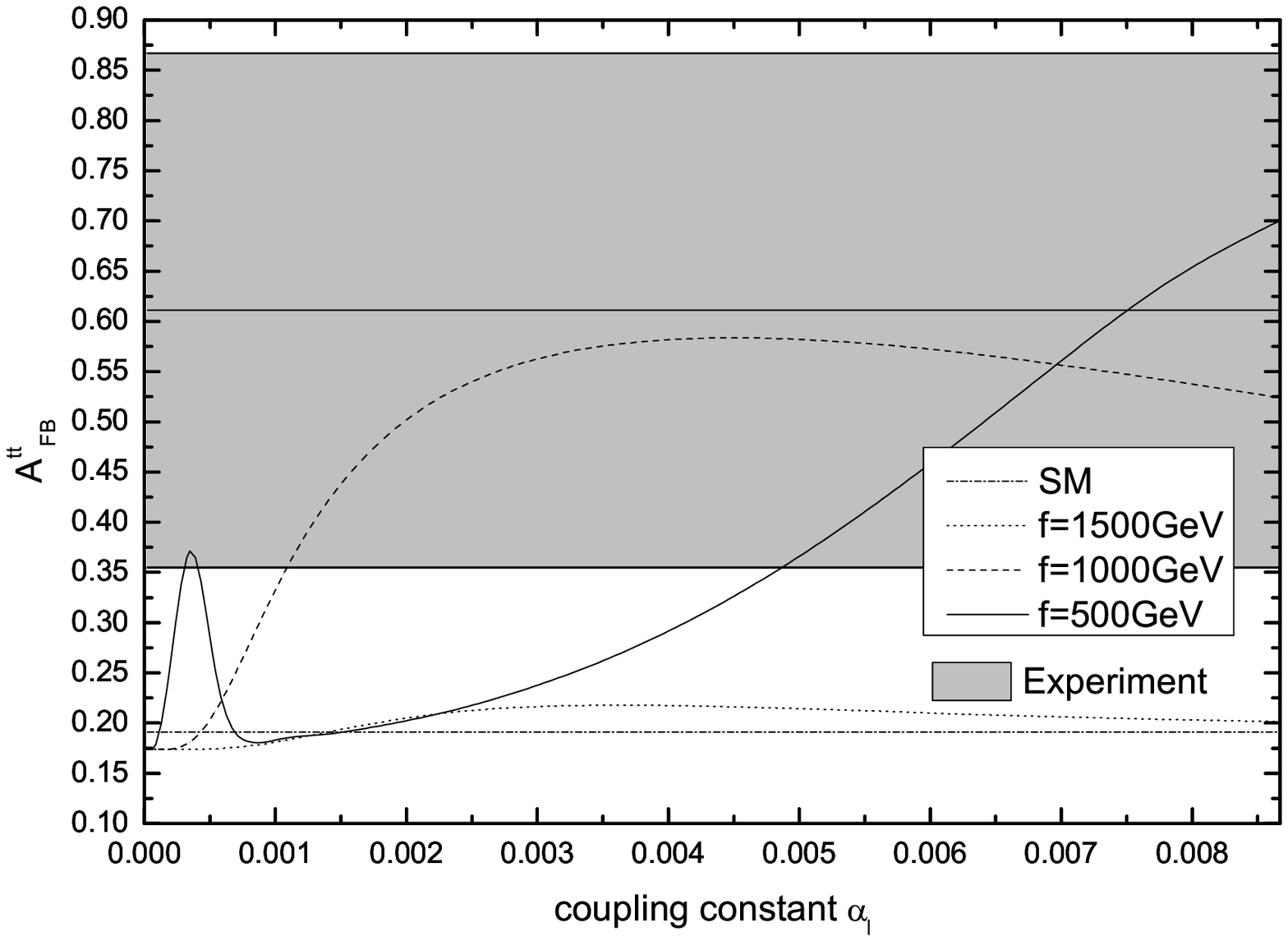}
\caption{The AFB with a cut of $M_{t\bar{t}}>450$GeV versus coupling constant in the frameworks of SM and LHM with $f=500$ GeV(solid), $f=1000$ GeV(dashed), $f=1500$ GeV(dotted).
The dash-dotted line stands for the values from SM\cite{Hollik:2011ps}, and the shadowed region is for experimental data\cite{Aaltonen:2011kc}.} \label{ycut}
\end{center}
\end{figure}

We also analyze our results for taking two different kinds of cuts. One is for $t\bar{t}$ $M_{t\bar{t}}>450\;GeV$ and the other is for
rapidity $|\bigtriangleup y|>1$. Those results are shown in Fig.\ref{mcut} and Fig.\ref{ycut}.

The Fig.\ref{mcut} shows that for $f=500$ GeV, as the coupling constant being near $0.0004$, the result is close to
the lower bound of the experimental data. While for $f=1000$GeV, the prediction fits the experimental value when the coupling constant is above $0.0018$.

Fig.\ref{ycut} shows that for $f=500$ GeV, taking the coupling constant within the range of $0.0003\sim0.0004$, the result fits experimental value.
For $f=1000$ GeV, the prediction fits the experimental value when the coupling constant is above $0.0012$.
From the above analysis we have reached some conclusions. First, for $f=500$ GeV, the prediction can coincide with experimental data when
the coupling constant takes the value about $0.0003\sim0.0004$. Second, for $f=1000$ GeV, the prediction could coincide with experimental data when
the coupling constant takes the value between $0.0018\sim0.0028$ and $0.0067\sim0.0087$.

%We also list the mass of $Z_H$ boson versus coupling constant in Fig.\ref{mzhmass}. For $f=500$GeV, the mass of $Z_H$ boson is around $450$GeV.
%For $f=1000$GeV, the mass of $Z_H$ boson is around $650\sim660$GeV and $750\sim850$GeV.

%\begin{figure}
%\setlength{\unitlength}{1mm}
%\begin{center}
%\includegraphics[width=12cm]{mzhmass.eps}
%\caption{The mass of $Z_H$ boson versus coupling constant with $f=500$GeV(solid), $f=1000$GeV(dashed), $f=1500$GeV(dotted).} \label{mzhmass}
%\end{center}
%\end{figure}

\section{Discussion and conclusion}

In this work, we study the contribution of LHM to the top quark forward backward asymmetry measured at Tevatron.
With the help of the software LoopTools, we calculate the tree diagram, box diagrams and their interference with the SM contributions.  We find that
only $Z_H$ boson in LHM makes a sizable contribution to the asymmetry when $f$ runs from $500$ GeV to $1000$ GeV.
As is understood, the new physics contribution should not significantly change the total cross section of the $t\bar t$ production, therefore there
exists a small parameter space which can reproduce the Tevatron asymmetry. When $f=500$ GeV, we predict that $Z_H$ boson should be of a mass
around $450$ GeV. While for $f=1000$ GeV, the mass of $Z_H$ boson is in two separate regions of $650\sim660$ GeV and $750\sim850$ GeV respectively. For the
expected ILC, whose center of mass energy is set to be $500$ GeV in early stage, we wish its the center of mass energy
may vary from $400$ GeV to $500$ GeV  in order to find whether the $Z_H$ boson with mass around $450$ GeV indeed exists.

\begin{acknowledgments}
This work is supported by the National Natural Science Foundation of China (10975027, 11275036, 11047002), the fund of the Natural Science Foundation of Hebei Province(A2013201277 , A2011201118) and Natural Science Fund of Hebei University (2011JQ05, 2007113).
the open project of State Key Laboratory of Mathematics-Mechanization(Y3KF311CJ1), and Natural Science Fund of Hebei University(2011JQ05, 2012-242).
\end{acknowledgments}

\end{document}